\documentclass{bmcart}

\usepackage[utf8]{inputenc} 

\usepackage{amsmath,amsfonts}
\usepackage{graphics}
\usepackage{graphicx}

\startlocaldefs

\newcommand{\bp}{\textbf p}
\newcommand{\thetas}{\boldsymbol{\theta}}
\newcommand{\etas}{\boldsymbol{\eta}}
\newcommand{\donotdisplay}[1]{}

\endlocaldefs

\begin{document}

\begin{frontmatter}

\begin{fmbox}
\dochead{Research}


\title{Complex decision-making strategies in a stock market experiment explained as the combination of few simple strategies}


\author[
   addressref={aff1,aff2},           
   corref={aff1,aff2,aff3},                       
   email={gael.poux-medard@univ-lyon2.fr\\marta.sales@urv.cat}   
]{\inits{G}\fnm{Gaël} \snm{Poux-M\'edard}}
\author[
   addressref={aff3},           
]{\inits{S}\fnm{Sergio} \snm{Cobo-Lopez}}
\author[
   addressref={aff4},           
]{\inits{J}\fnm{Jordi} \snm{Duch}}
\author[
   addressref={aff3,aff5},           
]{\inits{R}\fnm{Roger} \snm{Guimer\`a}}
\author[
   addressref={aff3},           
   email={marta.sales@urv.cat}           
]{\inits{M}\fnm{Marta} \snm{Sales-Pardo*}}


\address[id=aff1]{
  \orgname{Department of Physics, Ecole Normale Supérieure of Lyon}, 
  \postcode{69342},                                
  \city{Lyon},                              
  \cny{France}                                    
}
\address[id=aff2]{%
  \orgname{ERIC lab, Université de Lyon},
  \postcode{69361},
  \city{Lyon},
  \cny{France}
}
\address[id=aff3]{%
  \orgname{Department of Chemical Engineering, Universitat Rovira i Virgili},
  \postcode{43007},
  \city{Tarragona},
  \cny{Spain}
}
\address[id=aff4]{%
  \orgname{Department of Mathematics and Computer Science, Universitat Rovira i Virgili},
  \postcode{43007},
  \city{Tarragona},
  \cny{Spain}
}
\address[id=aff5]{%
  \orgname{ICREA},
  \postcode{08010},
  \city{Barcelona},
  \cny{Spain}
}


\begin{artnotes}
\end{artnotes}

\end{fmbox}


\begin{abstractbox}

\begin{abstract} 
Many studies have shown that there are regularities in the way human beings make decisions. However, our ability to obtain models that capture such regularities and can accurately predict unobserved decisions is still limited. We tackle this problem in the context of individuals who are given information relative to the evolution of market prices and asked to guess the direction of the market. We use a networks inference approach with stochastic block models (SBM) to find the  model and network representation that is most predictive of unobserved decisions. Our results suggest that users mostly use recent information (about the market and about their previous decisions) to guess. Furthermore, the analysis of SBM groups reveals a set of strategies used by players to process information  and make decisions that is analogous to behaviors observed in other contexts.
Our study provides and example on how to quantitatively explore human behavior strategies by representing decisions as networks and using rigorous inference and model-selection approaches. 
\end{abstract}




\end{abstractbox}
%

\end{frontmatter}



\section{Introduction}
In recent years, thanks to the widespread use of the internet, e-mail and mobile phone technologies, we have been able to gather large amounts of data that have enabled the large-scale characterization of specific traits of human behavior \cite{guimera03,kossinets06,gonzales08,candia08,schneider13}.
Indeed, a number of studies have shown that humans display statistically regular patterns in the way they move, communicate or make decisions which makes them identifiable \cite{ref4,ref5,ref6,saramaki14,godoy-lorite16a,ref38,montjoye13}. 
Despite the success in characterizing such systems, there has been, comparatively, less work to assess whether there are interpretable models of that are truly predictive of unobserved behavior \cite{ref38}.

A compelling example is that of the study of decisions made by individuals when playing dyadic games that represent social dilemmas. A recent study \cite{ref4} identified five different patterns in the strategies individuals use to play these games; these individual strategies, or behavioral phenotypes, deviate from optimal rational behavior and can be associated to common human attitudes such as jealousy, optimism or altruism \cite{ref4,ref7,ref15}. However,  such model is not the best for making predictions of unobserved individual decisions in the context of dyadic games. 
Indeed, a Bayesian modeling approach using a network representation of individual decisions in dyadic games showed that a model in which individuals mix three simple strategies is more predictive of unobserved individual decisions than a five-phenotype model \cite{cobo-lopez2018}.
Moreover, the Bayesian modeling approach reveals that the way individuals perceive games is different from the expectations based on game theory arguments \cite{cobo-lopez2018}.

Here, we use a similar network inference approach to model and understand  the decision-making process in the context of stock markets. Specifically, we consider the situation in which individuals have to guess the short-term direction ({\em up} or {\em down}) of a stock market based on real data given a reduced set of available sources of information, such as the evolution of the market in the previous time step, average historical market trends, information on other markets or the advice of an expert \cite{ref20}. Several questions arise in this context; for example: in an environment in which we are given more information than what we can process, which sources of information affect the decisions we make the most? How do we use this information to make decisions? Can we identify recurrent, context-dependent patterns of information usage that are predictive of unobserved decisions?

To address these questions, we assume that the information available to an individual fully defines the context in which she makes a decision; the decision is then the result of a specific strategy in which the individual exploits the available information. To define these strategies, we assume that there are some underlying patterns of behavior so that we can identify groups of individuals that display similar decision patterns and at the same time, identify groups of contexts that are similarly perceived by individuals. In our approach we exploit the fact that  we can represent the  decisions made by individuals in a specific context (for instance, the market went up in the previous round and the individual made a correct guess) as a bipartite graph connecting individuals and contexts, and use inference  techniques developed for complex networks. In particular, we use stochastic block models, a type of group-based generative models that are amenable to Bayesian inference methods, including rigorous model selection \cite{valles-catala18,cobo-lopez2018}. Our choice is further motivated by the fact that models in  which individuals can mix more than one strategy have been shown to be successful at predicting unobserved individual decisions in other contexts \cite{guimera12,godoy-lorite16a,cobo-lopez2018}.

With our network inference approach, we are able to rigorously compare network data representations that define contexts using different types of information---for instance information about the previous round or a historical trend. We find that users are rather Markovian when it comes to processing available information: individuals are best described if we assume that they only use information from the previous and current round to make their decisions. We also find that, consistently with some previous analysis \cite{ref20}, individuals use the information of the previous and current rounds in different ways to construct four distinct strategies: a switching strategy---individuals make a different decision at each time step; an optimistic strategy---individuals tend to predict that the market will go up; a repeating strategy---individuals repeat their previous decision; and a win-stay loose-shift strategy---individuals copy the previous market move.

Our inference approach thus makes it possible to identify the best representation of the data in terms of predictability. Additionally, through this representation we can explore the regularities in the way individual players use information to make decisions thus providing a valuable illustration of how power of inference methodologies to advance our understanding of human behavior from data.

\section{Methods}

\subsection{Dataset} 

Here, we consider the data set collected in the Mr. Banks social experiment \cite{ref20}\footnote{The datasets analysed during the current study are available in the Zenodo repository, https://zenodo.org/record/50429\#.YDOCwqvPxPY}. In this experiment, participants participated in a game that consisted in correctly predicting the evolution of a simulated market. That is, whether it would go up ($\uparrow$; stocks increase their value) or down ($\downarrow$; stocks lose value). In total, 280 people participated in this experiment. Participants played during 25 rounds that corresponded to 25 days of a real stock market. The experiment used 30 different time series of 25 days in a real market taken from the period between 01/02/2006-12/29/2009 of daily prices of: the Spanish IBEX, the German DAX and the S\&P500 from the United States.

In each round, players had access to the following sources of information: the evolution of the stock market during the previous month, a simulated expert's advice that, by design, was correct 60\% of the time, the trend of the same market in other places in the world, the average trend of the market over the previous 5 and 30 days, and the daily changes of direction of the market during the 30 previous days. We refer to this information as the {\em context} of the player.

\subsection{Network models and inference } 

\paragraph{Model selection} 
To find the best model for our data, we compare models in terms of their ability to predict unobserved decisions. Asymptotically, and most often in practice as well \cite{valles-catala18}, this is equivalent to using Bayesian inference for model selection. In the Bayesian setting, the best model is the one that has the largest posterior probability $p(M|D)$ (or equivalently the model that has the shortest description length \cite{valles-catala18,peixoto18}). Here, we maximize the posterior $p(M|D)$ to obtain model parameters, and follow a predictive approach to find, first, which type of SBM better describes observed data and, second,  which is the network representation with which we get the most accurate  predictions of unobserved decisions.

\paragraph{Network representation} Our goal is to predict whether a given player $p$ will guess that the market is going up or down when exposed to a specific context $c$, given a set of past observed decisions $R^o$ of the player herself and other individual players in different contexts. As noted above, a context $c$ is the  information accessible to  the player before making her decision.

We represent the data as a bipartite network in which nodes are players and contexts. We draw an edge with value 1 or 0 between a player $p$ and a context $c$ if, in that context, the player guessed $\uparrow$ or $\downarrow$, respectively. For instance, consider that the available information at round $t$ is: 1) whether the guess of the player at round $t-1$  was right or wrong, $C=\{{\rm R, W}\}$; and  2) the market evolution at time $t-1$, $B=\{\uparrow, \downarrow\}$. Then, there are four possible different contexts for all $CB$ combinations and we would represent our data as a bipartite network with $N$ players and four different contexts $c=\{({\rm R}\uparrow),({\rm W}\uparrow),({\rm R}\downarrow),({\rm W}\downarrow)\}$.

Because we have information about the guessing history of all users, it would be possible to build many bipartite graphs in which contexts consider only current information, current information and information from the previous round, current information and information from the previous two rounds, etc. However, for our inference approach to properly work, we need to keep the number of possible contexts small enough to detect statistical patterns between individuals and contexts. Besides that, we have {\it a priori} no justification to assume that some choices of context are better than others. In our approach, assessing the predictive power over unobserved decisions allows us to select both the best model and the best set of contexts to represent the data.

\paragraph{Single-membership and mixed-membership stochastic block models}
In our approach, we assume that there are statistical regularities in the way individuals make decisions in different contexts. These regularities define the strategic behaviors of players. In the network representation, we assume that the statistical regularities take the form of groups of players and groups of contexts with similar connections.

Consistent with this assumption, we model player decisions using stochastic block models \cite{white76,holland83,nowicki01,guimera09}. Stochastic block models are simple generative network models that assume the existence of groups of nodes (players and contexts), and that the probability that a pair of nodes connects (that is, that a player guesses up or down in a given context) depends exclusively on the groups to which the nodes belong.

Specifically, because the network we consider has two types of nodes, players and contexts, we consider bipartite stochastic block models \cite{guimera11,guimera12,yen20}. In addition, we consider two types of stochastic block models: one in which each player and context can belong to a single group (SBM), and another one in which players and contexts can belong to multiple groups simultaneously with different weights (mixed-membership SBM, MMSBM) \cite{ref12,ref16}. 

Formally, we have a set $U$ of players and a set $I$ of contexts, and the observed decisions $R^o = \lbrace r_{pc} \rbrace$ that players $p\in U$ make in context $c \in I$. In the data we consider that each player $p$ in a specific context $c$ has to guess the direction of the market in the next round, therefore the decision $d_{pc}$ is binary: $\uparrow$ or $\downarrow$.

We assume that there are $K$ groups of players and $L$ groups of contexts, and that the probability that player $p$ in group $k$ in context $c$ in group $\ell$ makes decision $d_{pc}=\uparrow$ is given by $p_{k\ell}$ where ${\bf p}$ is the matrix of connection probabilities between pairs of groups. Note that because there are only two possible decisions, the probability that $d_{pc}=\downarrow$ is $p_{k\ell}(\downarrow) = 1-p_{k\ell}$.

In the case of MMSBM, we allow players and contexts to belong to more than one group. We therefore introduce a membership vector for players $\thetas_{p}$, such that $\theta_{pk}$ is the probability that player $p$ belongs to group $k$. Analogously, we introduce a membership vector for contexts $\etas_c$ such that $\eta_{c\ell}$ is the probability that context $c$ belongs to group $\ell$. Because these vectors represent probabilities they are subject to the normalization conditions:
\begin{equation}
    \sum_k \theta_{pk}=1 \qquad \sum_{\ell} \eta_{c\ell}=1 ~.
    \label{eq.norm}
\end{equation}
Note that membership vectors become binary in  the single membership SBM model, so that player $p$ exclusively belongs to group $k$, $\theta_{pk}=1$ and $\theta_{pk'}=0$ for all  $k'\neq k$.

In the general MMSBM, the probability that player $p$ makes decision $d_{pc}$ in context $c$ is: 
\begin{eqnarray}
P(d_{pc}=\uparrow|\thetas,\etas,\bp) &=& \sum_k \sum_{\ell} \theta_{pk} p_{kl} \eta_{c\ell} \\
P(d_{pc}=\downarrow|\thetas,\etas,\bp) &=& \sum_k \sum_{\ell} \theta_{pk} (1-p_{kl}) \eta_{c\ell} ~.
\label{eq.prob}
\end{eqnarray}

SBMs further assume that each decision is independent from the others (conditionally on the group memberships), so that the probability of observing the data given the model parameters $\thetas, \etas,\bp$ (or likelihood) can be expressed as the product of the probabilities of each individual decision: 
\begin{equation}
  p(R^{\circ} \vert \thetas, \bp, \etas) = \prod_{(p, c) \in R^{\circ}} p(d_{pc}=\uparrow|\thetas,\etas,\bp)^{n^{\uparrow}_{pc}} p(d_{pc}=\downarrow|\thetas,\etas,\bp)^{n^{\downarrow}_{pc} } ~,
\end{equation}
where $n^{\uparrow}_{pc}$ is the  number of times player $p$ guesses $\uparrow$ in context $c$, and  $n^{\downarrow}_{pc}$ is the  number of times  player $p$ guesses $\downarrow$ in context $c$.

\paragraph{Inference and prediction}

A priori , we are agnostic about the values that the model parameters should take. We therefore use a non informative prior so that the  posterior probability of the model is proportional to the likelihood $p( \thetas, \bp, \etas | R^{\circ}) \propto p(R^{\circ} \vert \thetas, \bp, \etas)$. 
We then find the model parameters that maximize the posterior.
In the case of the  SBM we use simulated annealing to find the set of model parameters $(\thetas^*, \etas^*, \bp^*)$ that maximizes the posterior probability \cite{cobo-lopez2018}. In the case of the MMSBM, we use the expectation maximization approach described in \cite{ref9,cobo-lopez2018} (see \textit{Materials and Methods}). We make our predictions about unobserved decisions using these maximum a posteriori parameters.



\section{Results}
\subsection{Model selection}
We start by looking for the model that best describes our data, that is, the most predictive one. We consider three different models and assess their ability to make predictions measuring their accuracy on unobserved data using 5-fold cross validation \cite{valles-catala18}. The first model we consider is a {\em naive baseline}, in which we use the most common observed decision of player $p$ in context $c$ as a prediction. If there are no observed decisions of player $p$ in context $c$, we predict $d_{pc}=\uparrow$ because this is the most common decision (and the most common market move) in our data set. The other two models are the (single-membership) SBM and the (mixed-membership) MMSBM described above.

Each set of contexts defines a different network representation. We compute the average predictive accuracy of each model over folds and over network representations (Fig.~\ref{CompAcc}). We find that, overall, MMSBM models perform better than the naive baseline and single-membership SBMs. Therefore, in what follows we use MMSBMs to identify the best representation of the data and to analyze player strategies.

\subsection{Identification of the most predictive network representation of the data }

Next, we identify the most predictive network representation for our data so as to establish which pieces of information and mechanisms are being used in the decision-making process. As we have already mentioned, we have the full history of player's decisions, daily and average market evolution, and information on whether players consult the expert opinion before making their decisions. We find that past information beyond the previous round $t-1$ is not relevant to predict decisions at round $t$ (Fig.\ref{CompSit} and Supplementary Material).
We therefore consider 23 different network representations that mostly consider information about the user and the market at round $t-1$  and the expert's advice at round $t$, and compare the performance of MMSBMs (fit to each different network representations) at predicting unobserved data (Fig. \ref{CompSit}). To avoid fold-to-fold variability, we use the average log-ratio of the accuracy of pairs of models as our metric to compare predictive performance:
\begin{equation} 
Q_{S_1,S_2} = \frac{1}{N_{\rm folds}} \sum^{N_{\rm folds}}_{i=0} \log(\frac{A_{S_1, i}}{A_{S_2, i}}) ~,
\end{equation}
where $A_{S, i}$ is the predictive accuracy in fold $i$ for representation $S$, and $N_{\rm folds}$ is the total number of folds.  Note that by taking the logarithm of the ratio, we ensure that the metric is symmetric with respect to zero. 

We find that there are five representations that yield an almost identical accuracy, significantly higher than that of the other representations (Fig.~\ref{CompSit}). Among those, we select the simplest representation, which comprises 12 different contexts characterized by three sources of information available to players at round $t$: the market evolution at round $t-1$ ($\uparrow or \downarrow$), the outcome of the player's guess at round $t-1$ (right or wrong), and the expert's advice at round $t$ ($\uparrow$, not consulted, or $\downarrow$). Importantly, adding further information does not increase the predictive accuracy. Our analysis thus shows that players have short memory. This result correlates with the findings of Ref.~\cite{ref20} on the same data set. While there is a possibility that our data set is not big enough to capture effects beyond round $t-1$, our result is consistent with other studies which have successfully used a Markovian human in the analysis of  decision making processes \cite{McGhan2015,Lin2014,Karami2009}.

\subsection{Each group of users has well-defined patterns of behavior}

Next, we turn to the model parameters obtained for the most predictive representation of the data. In particular we focus on player groups, which we identify with guessing strategies \cite{cobo-lopez2018}. As mentioned before, MMSBMs assume that players can mix several strategies and that contexts can also belong to different groups simultaneously. In our case, we find that  the highest predictive accuracy is for a MMSBM with $K=4$ groups of players (or strategies) and $L=8$ groups of contexts (see Supplementary). Interestingly, we find that contexts tend to belong to a single group of contexts, while players have their memberships spread across different groups. In other words, the players behavior is the result of mixing different strategies (see Supplementary Material). 

The group-to-group probabilities $p_{k\ell}$ express the probability that a group of players $k$ guesses $\uparrow$ when facing contexts that belong to group $\ell$, and therefore encapsulate all the information of the strategy of each group of players.
For a more straightforward interpretation of the strategies (and because each context belongs mostly to a single group), we show the matrix $\hat{p}_{kc}=\sum_{\ell} p_{k\ell} \eta_{c\ell}$ corresponding to the probability that a group of players $k$ guesses $\uparrow$ in each context $c$ (Fig.~\ref{Pki}).
We refer to each one of the rows in $\hat{p}_{kc}$ as an elementary strategy, since players combine these elementary strategies to give rise to observed complex strategies.
Figure ~\ref{Pki} shows that elementary strategies are well defined because the $\hat{p}_{kc}$ are often either close to zero or one. 

To help summarize and interpret elementary strategies, we note that in the $\hat{p}_{kc}$ matrix we can identify several simple, easily interpretable patterns that we can use as alternative building blocks to describe decision-making strategies:
\begin{itemize}
\item Win-stay (WS):  at round $t$ repeat the guess of round $t-1$ if it was right.
\item Lose-shift (LS): at reoun d$t$ change the guess from round $t$ if it was wrong at $t-1$.
\item Market imitation (MI): the guess at round $t$ matches the market direction at $t-1$. This strategy is equivalent to a combination of the two previous strategies which we call WSLS: if the market goes $\uparrow$ the player guessed $\uparrow$, MI=WS; if the player guessed $\downarrow$, then MI=LS. 
\item Repeat previous guess (RPT): the guess at $t$ is equal to the  guess at $t-1$.
\item Repeat after $\uparrow$ (RPTU) (or $\downarrow$, RPTD): repeat last guess if it was $\uparrow$ (respectively, $\downarrow$).
\item Expert's advice (EXP): follow the expert's advice if consulted.
\end{itemize}
The first three of patterns were already identified in the behavior of some players in the original data collection study, as well as a general bias towards $\uparrow$ when making decisions \cite{ref20}. However, whether these strategies were combined with others or used by different users was not explored.

To assess how each one of the elementary strategies aligns with these building blocks for behavioral patterns, we define for each group $k$ a score $M_{kb}$ that quantifies the extent to which players within group $k$ follow behavioral pattern $b \in \{{\rm MI, WS, LS, RPT, RPTU, RPTD, EXP }\}$:
\begin{equation}
M_{kb} = \frac{1}{N_{kb}}\sum_{c \in S_b}N_{kc}\,\big[q_{kc}(b)-(1-q_{kc}(b))\big] ~.
\end{equation}
In the summation, $S_b$ represents the set of contexts in which  a behavioral pattern can be observed -- for example, while RPTD can be observed after a user guessed down in the previous round, RPTU cannot --, $q_{kc}(b)$ is the probability that a user in group $k$ follows the elementary strategy $b$ in context $c$, $N_{kc}$ is the number of times a player in group $k$ faces context $c$, and $N_{kb}=\sum_{c\in s_b} N_{kc}$ is the total number of observations of players in group $k$ facing contexts $c \in S_b$. 
Therefore, a score $M_{kb}=1$ means that the group $k$ always follows a pattern of behavior $b$ when facing contexts $c \in s_b$;  a score $M_{kb}=-1$ means that the group does exactly the opposite of what the strategy prescribes in 100\% of cases. For instance, if group $k$ always guesses $\uparrow$ regardless of the context, the scores for RPTU and RPTD will be $M_{k\,{\rm RPTU}}=1$ (always guess $\uparrow$ after having guessed $\downarrow$) and $M_{k\,{\rm RPTU}}=-1$ (always guess $\uparrow$ after having guessed $\downarrow$).

In Fig.~\ref{InfStrat}, we show the $M_{kb}$ scores for each of the groups of players. First, we note that when the expert is consulted, three of the  groups follow her advice, while the remaining group is not influenced by it. 
When the expert is not consulted, we find that each group follows a very specific strategy. Group $k_1$ strongly tends to change their previous decision (switching behavior); $k_2$ tends to guess $\uparrow$ (optimistic behavior); $k_3$ almost always repeats the previous decision (repeating behavior); and $k_4$ always follows a win-stay loose-shift (WS-LS) behavior.

\section{Discussion}

Our analysis shows that the best description of player's strategies to predict their decision about market evolution is to consider that players are mixing elementary strategies which we have termed switching, optimistic, repeating, and win-stay (WS-LS).

Interestingly, not all of these elementary strategies are combined in the same way by all of the players (Fig.~\ref{mainPlot}). The most used strategy overall is the WS-LS strategy, which on average explains 34\% of player's decisions, that is, $\frac{1}{N}\sum_u \theta_{u,{\rm WS-LS}} = 0.34$. We observe this trend for both  players using a single elementary strategy (low entropy in Fig.~\ref{mainPlot}) and those combining various elementary strategies (high entropy). This finding is consistent with the conclusions of numerous studies assessing the wide use of this behavior in various natural systems such as children education \cite{ref23,ref31}, evolutionary systems \cite{ref26,ref27}, or games involving adversity \cite{ref25}.

By contrast, the optimistic behavior is seldom used by users using a single strategy (Fig.~\ref{mainPlot}a, d). Nonetheless the optimistic strategy is on average part of 25\% of payers' behavior, suggesting that this is a common strategy when used in combination with other strategies, which reflects the overall bias to guess $\uparrow$ regardless of the context in which the decision is made \cite{ref20}.

Finally, we note that that the two remaining elementary behaviors (switching decisions and repeating the previous decision) are complementary to one another, since if a player does not repeat her decision, it means that she changes her decision. Therefore, a player combining both  strategies with equal probability would generate a sequence of random guesses. To investigate whether these strategies are the result of a behavioral pattern or just capture random sequences of $\uparrow$ and $\downarrow$, we measure from each player $p$, the normalized probability difference of use of each one of the two strategies $D_p$:
\begin{equation}
  D_p = \frac{\theta_{p\,{\rm RPT}} - \theta_{p\,{\rm SWI}}}{\theta_{p\,{\rm RPT}} + \theta_{p\,{\rm SWI}}}  ~.
\end{equation}
If $D_p=1$  player $p$ uses exclusively the repeating strategy; if $D_p=-1$ means she exclusively uses the shifting strategy; if  $D_p=0$ means that both strategies are used equally. We compute $D_p$ only for players that use either "shifting" or "repeating" as a main strategy to avoid considering players that mostly use other strategies (WS-LS or optimistic). We find that while there are two peaks at extreme values ($D_p=-1$ and $D_p=1$) showing that many players use exclusively one of the two strategies, players have a tendency to repeat their previous guess (i.e. the distribution of $D_p$ is skewed towards $D_p=1$).  Indeed, this observation shows that some players behave according a natural persistence cognitive bias observed when an individual has to make repeated decisions \cite{persistence,persistence2}.

\section{Conclusion}

In this work we have shown the suitability of the MMSBM for the study of social systems, particularly for modeling, predicting and understanding human decision-making processes. MMSBMs not only provide more accurate predictions than other state-of-the-art methods \cite{cobo-lopez2018,guimera12}, but they are interpretable.

A close analysis of the model parameters highlights the  ability of these models to decompose complex behaviors into a linear combination of elementary behaviors. Our analysis helps identify basic patterns of behavior that are consistent with human behavior in other contexts. First, we observe an optimistic bias that suggests that individuals make decisions based on their desire that the market goes up. Second, we observe that many players use a win-stay loose-shift strategy which is known to be very efficient in the long run in learning processes, games of adversity, co-evolution networks, etc. Last, we find that players also display a tendency to repeat previous decisions, a behavior observed when individuals have to perform repetitive tasks. Our study demonstrates that these behavioral patterns of behavior are hidden to the naked eye but can be obtained from the data using our approach. 

MMSBM have been widely used in several other fields including quantitative and computational social science \cite{cobo-lopez2018}. This work confirms and pinpoints its relevance in the study of social systems. Its predictive accuracy and interpretability could be efficiently exploited in different studies and experiments in sociology and psychology in order to improve the understanding of human behavior \cite{hofman17}. Moreover, MMSBM offer an alternative framework to conventional populations analysis tools used in social sciences, and therefore finds its direct uses in the confirmation and information of state-of-the-art theories in cognitive sciences, sociology, psychology, etc., if not for the discovery of previously unreachable hidden thought mechanisms.

\section*{Materials and methods} \hfill

In order to maximize the logarithm of the likelihood, we use a variational approach. First we use a trick to change the logarithm of a sum into a sum of logarithms :

\begin{equation} \label{eq1}
\begin{split}
log(P(R^{\circ}& \vert \theta, \eta, p)) = \sum_{(ui)\in R^{\circ}} N_{r_{ui}}\log(\sum_{k,l} \theta_{uk} p_{kl}(r_{ui}) \eta_{il}) \\
&= \sum_{(ui)\in R^{\circ}} N_{r_{ui}} \ln(\sum^{K, L}_{k, l} \omega_{ui}(k,l) \frac{\theta_{uk} p_{kl}(r_{ui}) \eta_{il}}{\omega_{ui}(k,l)})\\
&\geq \sum_{(ui)\in R^{\circ}} N_{r_{ui}} \sum^{K, L}_{k, l} \omega_{ui}(k,l) \ln(\frac{\theta_{uk} p_{kl}(r_{ui}) \eta_{il}}{\omega_{ui}(k,l)})
\end{split}
\end{equation}

In the first line of Eq.\ref{eq1}, we introduced the $\omega_{ui}(k,l)$ function, which is the probability for a node u belonging to the group k to be linked by an edge $r_{ui}$ to the node i belonging to the group l. Using Jensen's inequality $\ln(\bar{x})\geq \overline{ \ln(x) }$, we end up to the last line of the equation \ref{eq1}. Then, one notices that this inequality becomes an equality for:

\begin{equation} \label{eq2}
\omega_{ui}(k, l) = \frac{\theta_{uk} \cdot p_{kl}(r_{ui}) \cdot \eta_{il}}{\sum^{K, L}_{k', l'}\theta_{uk'} \cdot p_{k'l'}(r_{ui}) \cdot \eta_{il'}}
\end{equation}

Which is the update equation for the expectation step. In order to maximize the log-likelihood, we derive it with respect to $\theta$, $\eta$ and p using Lagrange's multipliers to account for normalization constraints. We obtain:

\begin{equation}
\theta_{uk} = \frac{\sum_{r_{ui}} N_{r_{ui}} \sum_{i \in \partial u} \sum^{L}_{l} \omega_{ui}(k,l)}{d_{u}}
\end{equation}
\begin{equation}
\eta_{il} = \frac{\sum_{r_{ui}} N_{r_{ui}} \sum_{u \in \partial i} \sum^{K}_{k} \omega_{ui}(k,l)}{d_{i}}
\end{equation}
\begin{equation}
p_{kl}(r) = \frac{\sum_{r_{ui}} N_{r_{ui}} \sum_{(u,i) \in R^{\circ} | r_{ui}=r} \omega_{ui}(k,l)}{\sum_{(u,i) \in R^{\circ}} \omega_{ui}(k,l)}
\end{equation}

With $d_{u}$ the degree of node u and $d_{i}$ the degree of node i.


\begin{backmatter}

\section*{Competing interests}
  The authors declare that they have no competing interests.

\section*{Author's contributions}
    GPM performed technical work, interpreted the results, wrote the article. SCL helped the technical work, interpreted the results, wrote the article. JD gathered the dataset. RG interpreted the results. MSP interpreted the results, wrote the article.

\section*{Acknowledgements}
  R.G., J.D. and M.S.-P. acknowledge the support of MINECO through awards  FIS2016-78904-C3-1-P and  PID2019-106811GB-C31.
  

\bibliographystyle{bmc-mathphys} 
\bibliography{_Bibliographie}      




\section*{Figures}

\begin{figure}[h!]
\label{CompAcc}
\centering
    \includegraphics[width=.9\textwidth]{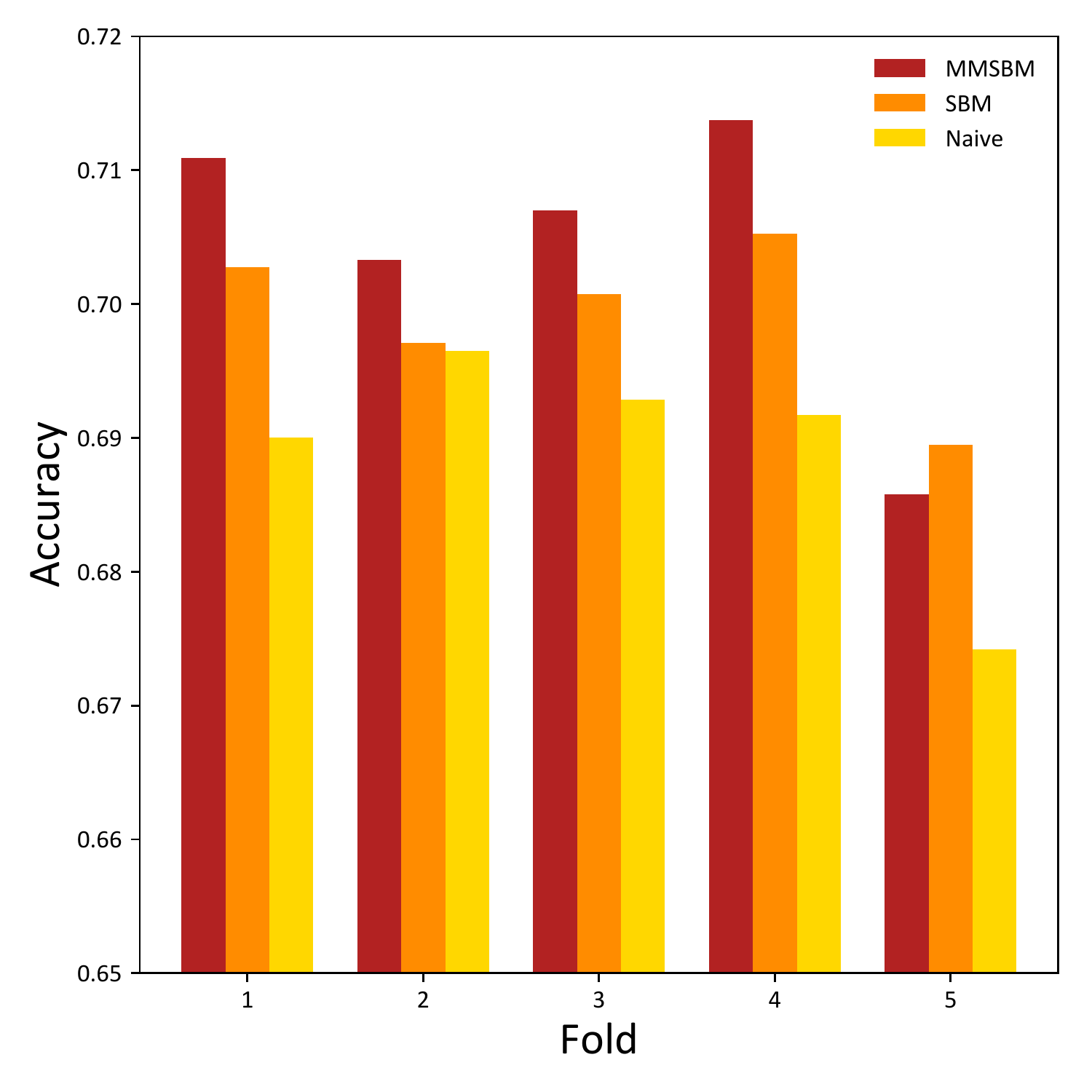}
    \caption{\csentence{Predictive accuracy of the three models}
    For each fold of the data, we show the predictive accuracy of the three models: naive, single-membership SBM and MMSBM. Each bar represents the average over 23 different network representations (as detailed in Figure~\ref{CompSit}), which combine the variables that are more informative of players decisions (player's decision at $t-1$, market evolution at $t-1$, outcome of decision at $t-1$, expert consultation, advice consulted, average market trend over the 5 last/all rounds, market's evolution/outcome before the previous round; see Supplementary Material).}
\end{figure} 

\begin{figure}[h!]
\label{CompSit}
\centering
    \includegraphics[width=.9\textwidth]{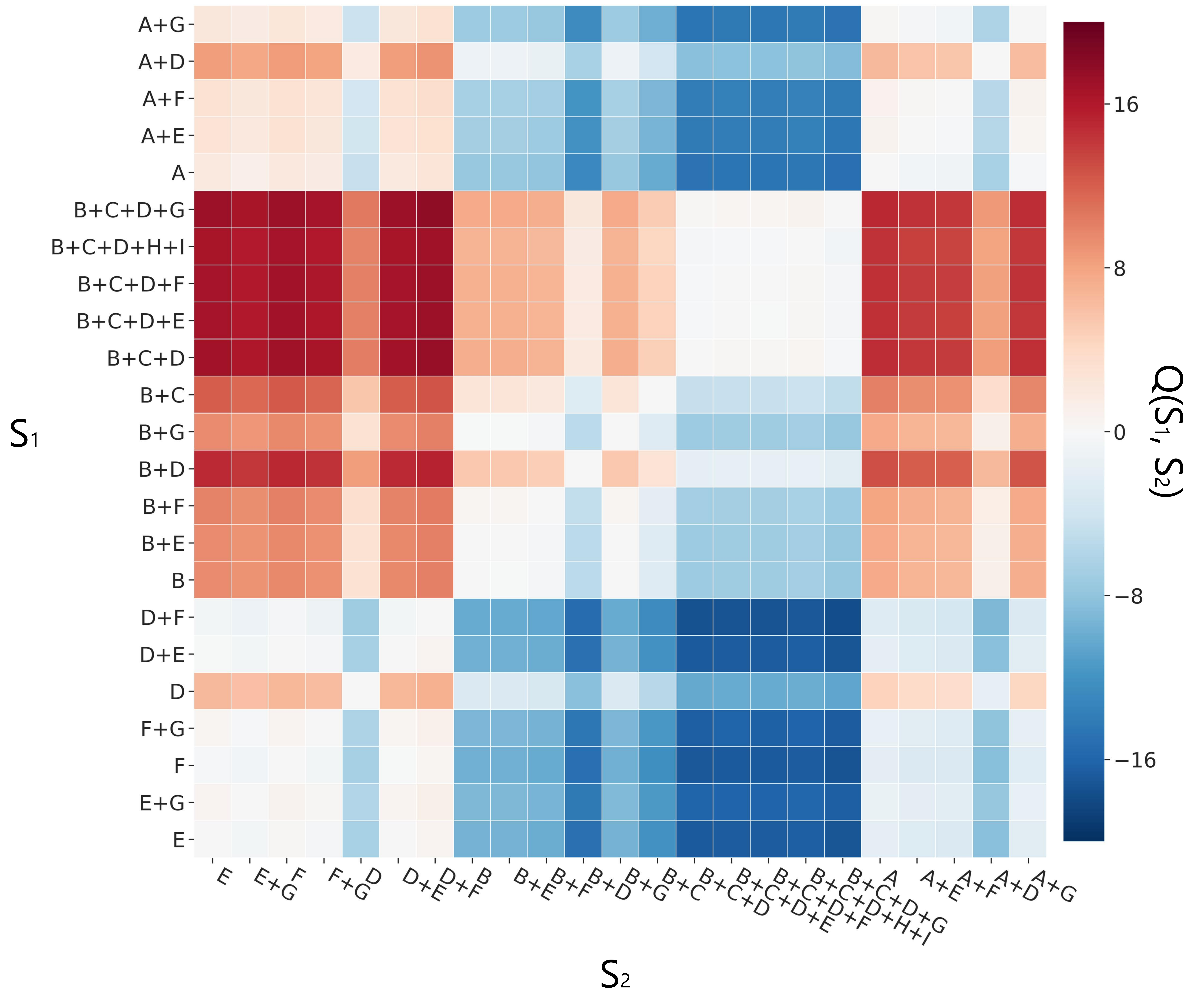}
    \caption{\csentence{Comparison matrix of data representations using predictive accuracy}
        Each row/column corresponds to a specific data representation $S$. We label them according to the information we are using to define the set of contexts: A=player's decision at $t-1$, B=market evolution at $t-1$, C=outcome of decision at $t-1$, D=expert consultation, E=indications consulted, F/G=average over the 5 last/all rounds, H/I=market's evolution/outcome before the previous round.
        Each matrix element  $Q_{S_1S_2}$ corresponds to the average log ratio of predictive accuracies (see text)  between representations $S_1,S_2$ and is colored following the colorbar on the right hand side. Note that if  $Q_{S_1S_2}>1$ (red), $S_1$(row) has a larger predictive accuracy than $S_2$ (column). If $Q_{S_1S_2}<1$ (blue), $S_1$(row) has a lower predictive accuracy than $S_2$ (column).}
\end{figure}

\begin{figure}[h]
\label{Pki}
\centering
    \includegraphics[width=.9\textwidth]{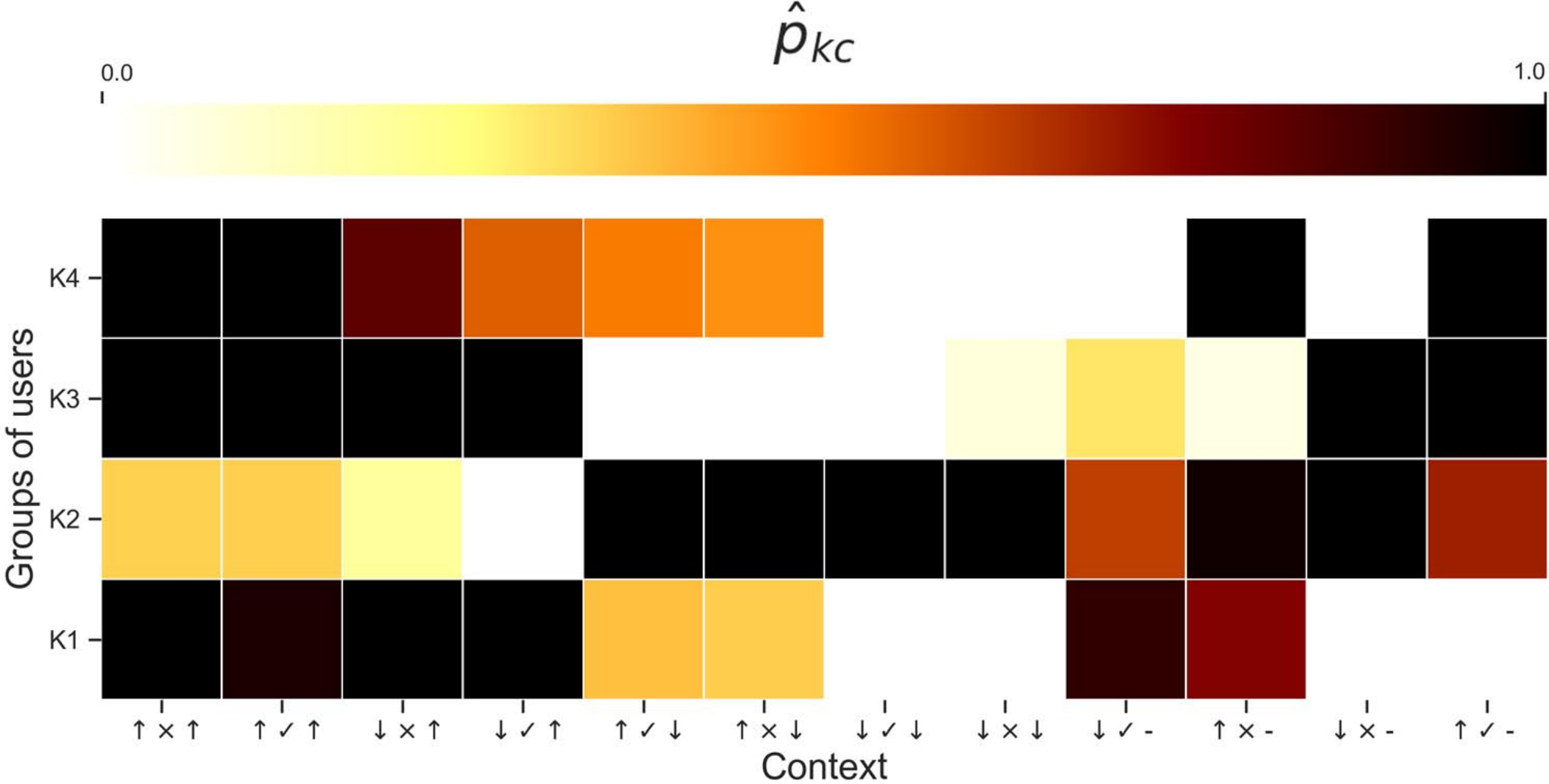}
    \caption{\csentence{Inferred probability matrix for a group of users to guess $\uparrow$}
    The symbols on the x-axis represent the context according to the following notation: $\text{market move} \in \lbrace\uparrow,\downarrow\rbrace$, $\text{outcome of previous guess} \in \lbrace \text{right}\checkmark$, $\text{wrong}\times \rbrace, \text{expert's guess} \in \lbrace \uparrow, \downarrow, \text{-} \rbrace$. We see the groups tend to describe well defined behaviors since the majority of matrix elements are either 1 or 0.}
\end{figure}

\begin{figure}[h]
\label{InfStrat}
\centering
    \includegraphics[width=.9\textwidth]{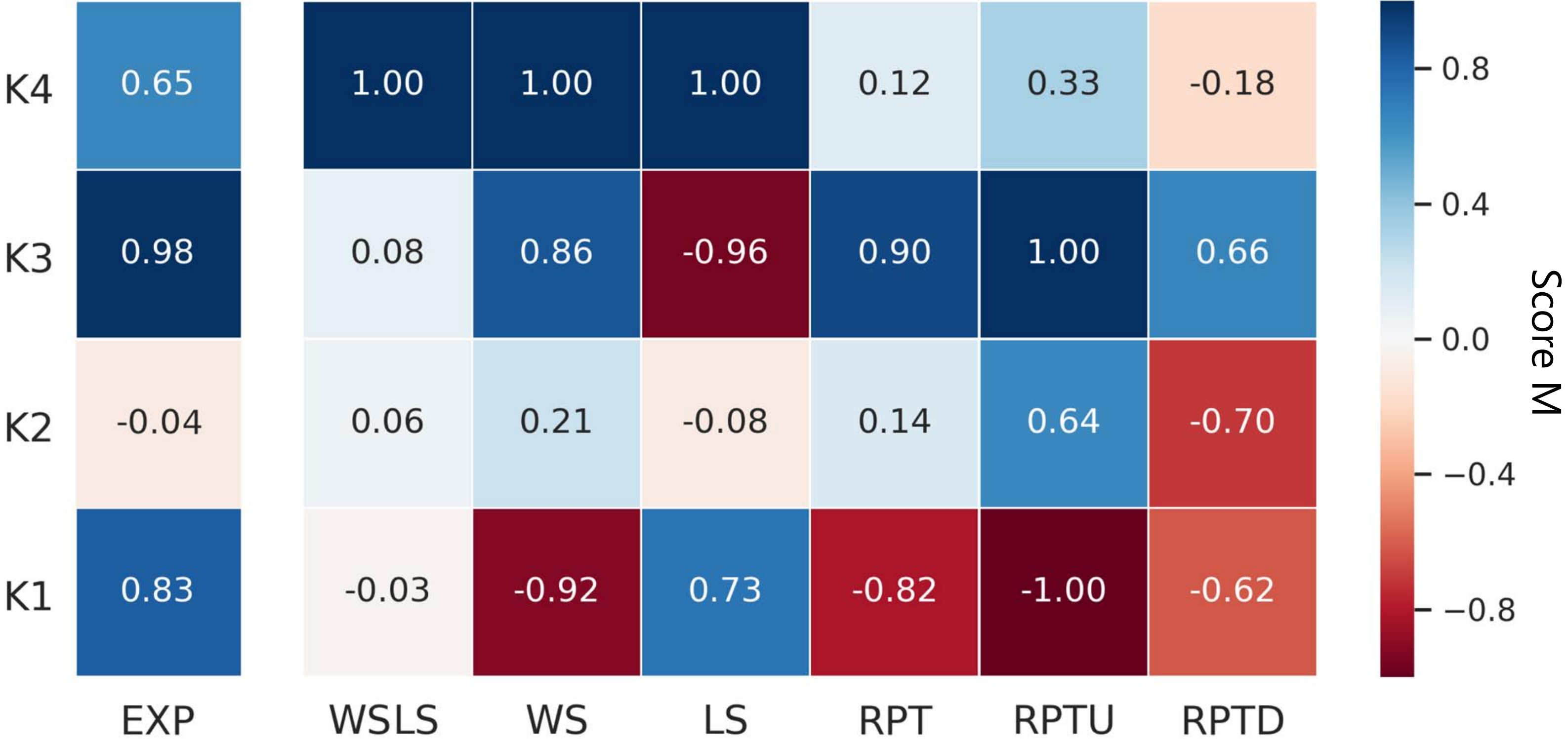}
    \caption{\csentence{Patterns of behavior for each group of users}
    For each group of users we show the score $M_{kb}$ (see text) for each pattern of behavior $b$ (from left to right): (EXP) Following the expert's advice; (WSLS) - Win-stay loose-shift; (WS) Win stay; (LS) Loose shift, (RPT) Repeat previous guess; (RPTU) Repeat previous decision if it was $\uparrow$; RPTD Repeat previous decision if it was $\downarrow$. A score of 1 means the group completely follows a basic pattern, and a score of -1 means it does the exact opposite of the basic pattern. Every group displays a preferential (if not systematic) pattern of behavior: K4 behaves according to the WS-LS strategy, K3 repeats the previous guess, K2 always guesses $\uparrow$ and K1 changes the previous guess. Additionally, the expert's advice is either followed or ignored, but none of the groups systematically makes the opposite decision.
    Note that because the expert is consulted in less than 30\% of the rounds, we split the data set into expert consulted and not consulted. From the former we get the score for the EXP strategy, while from the latter we obtain the scores for the other behavioral patterns.}
\end{figure}

\begin{figure*}[h!]
\label{mainPlot}
\centering
    \includegraphics[width=.9\textwidth]{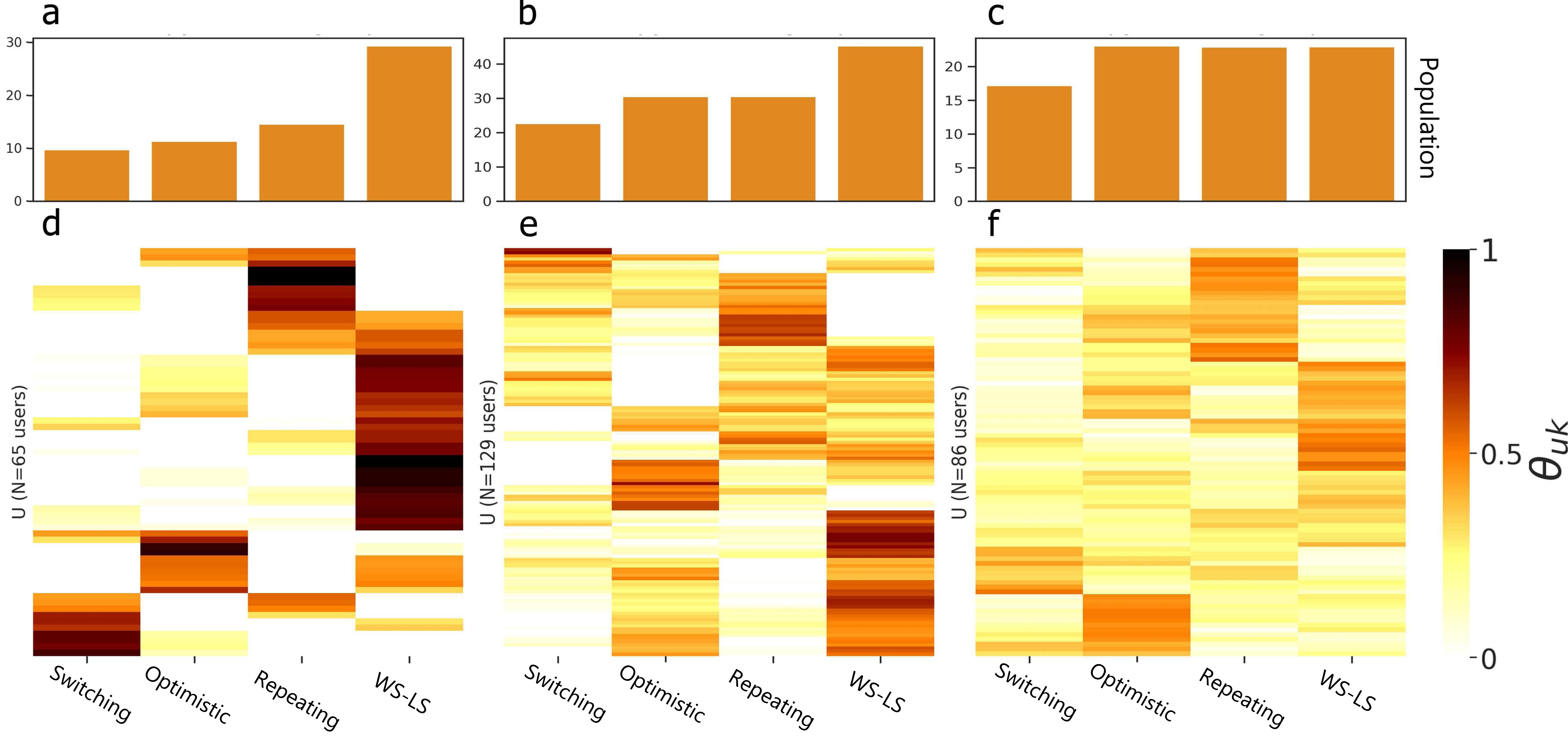}
    \caption{\csentence{Distribution of use of behavioral patterns by players}
        We compute the Shannon entropy of the $\theta$ membership vectors among four behavioral patterns: switching, optimistic, repeating, and win-stay (WS-LS). We split the users according to their Shannon entropy: ({\bf a,d}) Low entropy, including players who typically do not mix elementary strategies; ({\bf b,e}) Medium entropy, including players who mix two or three elementary strategies; ({\bf c,f}) High entropy, including players mix  three or four elementary strategies. ({\bf a,b,c}) Use of each elementary strategy for players with membership vectors with low, medium and high entropy users, respectively. {\bf (d,e,f)} Membership vectors. Each row represents one player. The color code quantifies the extent to which a player uses a behavioral pattern. We observe that the win-stay loose-Shift strategy is widely used alone, and that optimistic is commonly used in combination with other behavioral patterns.}
\end{figure*}

\begin{figure}[h]
\label{DistShiftStay}
\centering
    \includegraphics[width=.9\textwidth]{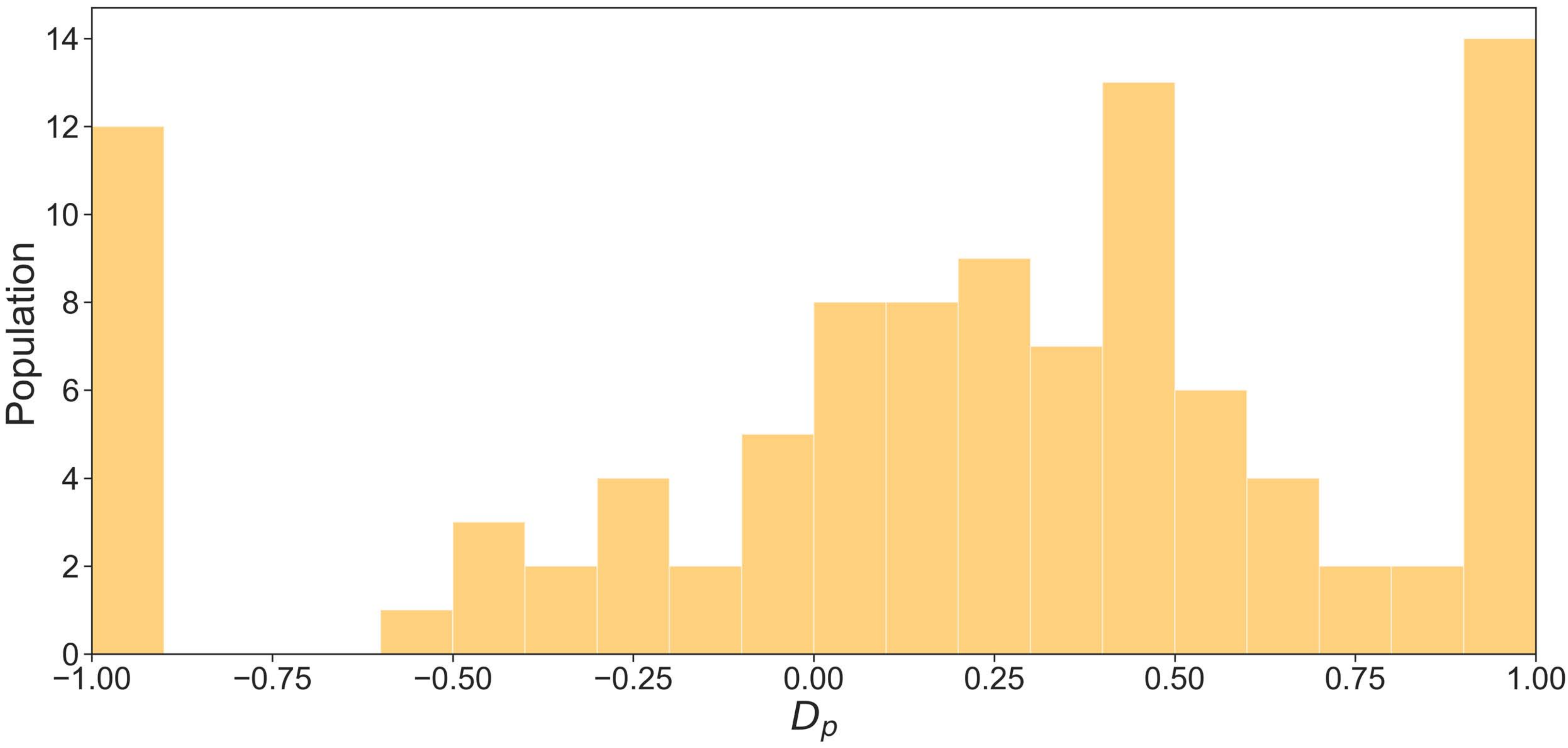}
    \caption{\csentence{Balance between repeat and switch strategies} We define $D_p = \frac{\theta_{p\,{\rm RPT}} - \theta_{p\,{\rm SWI}}}{\theta_{p\,{\rm RPT}} - \theta_{p\,{\rm SWI}}}$, so that
        $D_p$=0 means that both strategies are used by players to the same extent. This plot shows that a significant number of players are using exclusively one of the strategies; other players use a combination of those with a bias towards repeating their previous guess, rather than switching.}
\end{figure}

\clearpage


\section*{Additional Files}
  \subsection*{Additional file 1 --- Supplementary for: Complex decision-making strategies in a stock market experiment explained as the combination of few simple strategies}
    The file is in a PDF format.

\end{backmatter}
\end{document}